\documentclass[fleqn,usenatbib]{mnras}
\newcommand{\be}{\begin{equation}}
\newcommand{\ee}{\end{equation}}
\newcommand{\bea}{\begin{eqnarray}}
\newcommand{\eea}{\end{eqnarray}}
\usepackage{newtxtext,newtxmath}
\usepackage{rotating}

\usepackage[T1]{fontenc}

\DeclareRobustCommand{\VAN}[3]{#2}
\let\VANthebibliography\thebibliography
\def\thebibliography{\DeclareRobustCommand{\VAN}[3]{##3}\VANthebibliography}

\usepackage{graphicx}	
\usepackage{amsmath}	

\title[Deep learning large scale reconstruction]{Deep learning reconstruction of the large scale structure of the Universe from luminosity distance}

\author[C. Garcia et al.]{
Cristhian García,$^{1}$
Camilo Santa,$^{1}$
Antonio Enea Romano$^{1,2}$
\\
$^{1}$ICRANet, Piazza della Repubblica 10,I–65122 Pescara, Italy\\
${}^{2}$Instituto de Fisica, Universidad de Antioquia, A.A.1226, Medellin, Colombia \\
}

\pubyear{2022}

\begin{document}
\label{firstpage}
\pagerange{\pageref{firstpage}--\pageref{lastpage}}
\maketitle

\begin{abstract}
Supernovae Ia (SNe)  can provide a unique window on the large scale structure (LSS) of the Universe at redshifts where few other observations are available, by solving the inversion problem (IP) consisting in  reconstructing the LSS from its effects on the observed luminosity distance. So far the IP was solved assuming some restrictions about  space-time, such as spherical symmetry for example, while we obtain for the first time solutions of the IP problem for  arbitrary space-time geometries using  deep learning.
The method is based on the use of convolutional neural networks (CNN) trained on simulated data. The training data set is obtained by first generating random density and velocity fields, and then computing their effects on the luminosity distance. The CNN, based on an appriately modified version of U-Net to account for the tridimensionality of the data, is then trained to reconstruct the density and velocity fields from the luminosity distance.

We find that the velocity field inversion is more accurate than the density field, because the effects of the velocity on the luminosity distance only depend on the source velocity, while in the case of the density it is an integrated effect along the line of sight, giving rise to more  degeneracy in the solution of the IP. Improved versions of these neural networks, modified to accommodate the non uniform distribution of the SNe, can be applied to observational data to reconstruct the large scale structure of the Universe  at redshifts at which few other observations are available.  
\end{abstract}




\section{Introduction}
Supernovae are among the brightest astrophysical objects which can be observed in our Universe, and this has allowed to test the standard cosmological models at redshifts at which few other observations are available, and have provided the first evidence of an accelerating expansion \cite{Riess_1998}. 
Beside being useful to determine the background cosmological model parameters, they can also be used to reconstruct the density and peculiar velocity fields \cite{Odderskov_2017}, providing a unique tool to probe large scale structure at scales where other astrophysical objects are too dim to be observed. In the context of the luminosity distance the IP has been only solved assuming spherical symmetry \cite{Romano_2014}, and it consisted of solving complicated systems of differential equations, which required smooth functions as inputs, but observational data is rarely in a smooth form, limiting the accuracy of the results.
In this paper we will develop a completely new inversion method, which does not assume any symmetry.

The inversion  problem is a very general subject of investigation, studied in many different fields such as medical physics \cite{Guasch2020} or seismic inversion \cite{zhang2019datadriven}, and recently deep learning has shown to be a promising approach \cite{senouf2019selfsupervised,henzler2018singleimage,chen2020seismic,zhu2020general} for its solution, taking advantage of the computational advances made possible by the availability of graphical processing units (GPU). 
The method we adopt is based on creating a database of simulated luminosity distance data obtained by solving the direct problem for a large set or random density and velocity configurations. The simulated data is then used to train a convolutional neural network (CNN) to solve the IP, i.e. to reconstruct the density and velocity fields from the luminosity distance.
Since the physics of the direct problem is well understood \cite{Bonvin_2006}, there is virtually no limit in the amount of simulated data which can be created to train the CNN, allowing to obtain good results in the learning process, since deep learning performs well when a large and good quality training dataset is available.
We first generate random density fields using the nbodykit package \cite{Hand:2017pqn}, use these as inputs for solving the direct problem, and finally use the results of the direct problem as a training data set for the CNN.

This paper is organized as follows: the second section defines the IP, the third section explains how the mock density and velocity fields are generated, the fourth section shows how the effects on the luminosity distance are calculated from the density and velocity fields, the fifth section describes the neural network architecture we adopt, the sixth session defines the cost function which is minimized and provide details about the training process, the seventh and eighth and ninth sessions show the performance of the CNN in reconstructing the density and velocity fields. The last three sessions discuss the results of the inversion, how the solution of the IP could be improved with new CNN architectures, and how the method could be applied to observational data.

\section{Inversion problem}
Previous attempts to solve the IP were based on assuming spherical symmetry \cite{Romano:2013bxa,Chiang:2017yrq,Romano:2016utn} and solving a complicated set of differential equations.
The differential equations to be solved in these methods require as input a smooth function for the luminosity distance $D_L(z)$, implying the need of pre-processing the data by performing some fit.
In order to overcame the above limitations on the space geometry and the input data, we consider a completely general definition of the IP, consisting in reconstructing the density contrast and velocity fields from its effects on the luminosity distance using a CNN.

The CNN is trained with $10^3$  random density and velocity profiles generated  using the nbodykit package  \cite{Hand:2017pqn}. The CNN training set is then obtained by computing the effects of these inhomogeneities on the luminosity distance. Finally the IP is solved by training the CNN to reconstruct the density and velocity fields from the luminosity distance synthetic data obtained in the previous step.

\section{Simulation of cosmic structure}
In order to train the neural network we need realistic cosmic structure simulations, used to compute the effects of inhomogeneities on the luminosity distance. 
Since we need many independent mock density and velocity fields  to create a sufficiently large training and test set, N-body simulations would be very computationally expensive, and for this reason we have resorted to the nbodykit package \cite{Hand:2017pqn}. 
\begin{figure}
    \centering
    \includegraphics[scale=0.32]{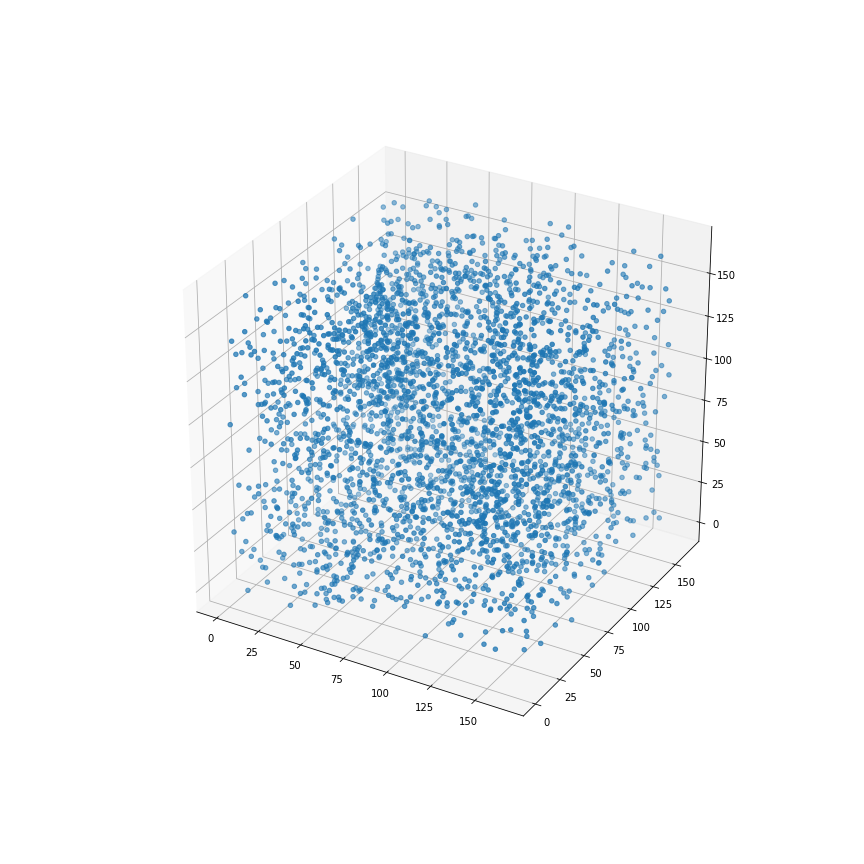}
    \caption{Example of random density field obtained using the LogNormalCatalog functionality of the package nbodykit, for a cube of size 300 Mpc. The observer is located at the center of the cube. In order to create sufficiently large training and test data sets $10^3$ independent mock fields realizations were generated.}
    \label{df}
\end{figure}

\begin{figure}
    \centering
    \includegraphics[scale=0.32]{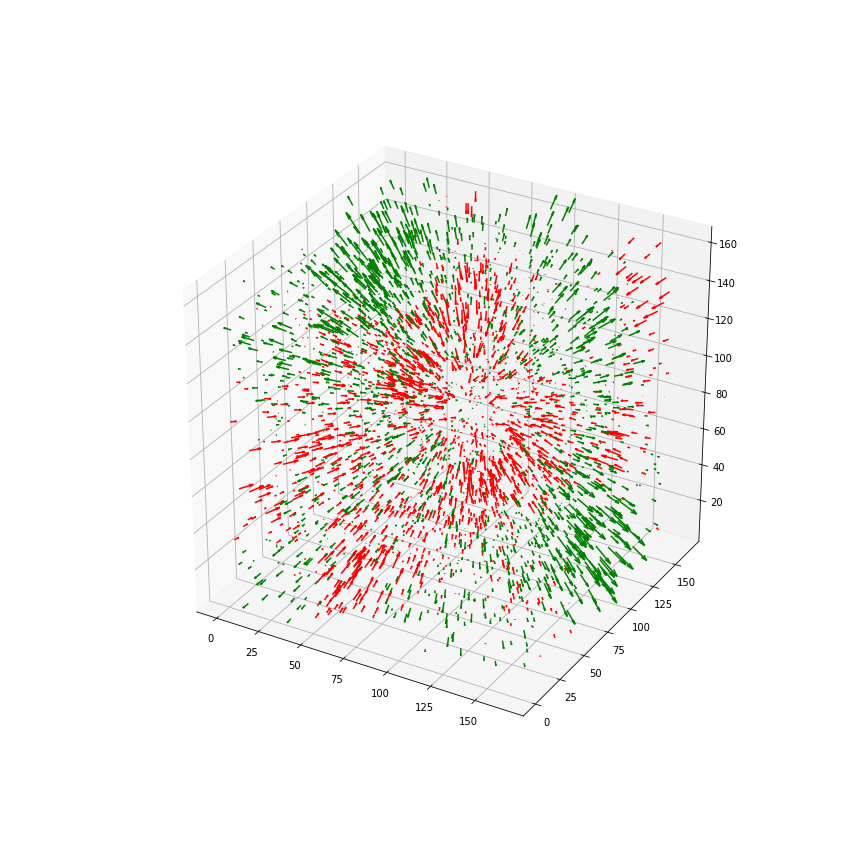}
    \caption{Example of random radial velocity field obtained using the LogNormalCatalog functionality of the package nbodykit, for a cube of size 300 Mpc. The observer is located at the center of the cube. Inward and outward directed velocities vectors correspond respectively to green and red arrows. For each simulation the velocity field is obtained from the corresponding density field by solving the Euler equations.}
    \label{vf}
\end{figure}
In particular we have used  the LogNormalCatalog functionality, which  generates a set of objects by Poisson sampling a log-normal density field and applies the
Zeldovich approximation to model nonlinear evolution. For each simulation the velocity field is obtained from the density field by solving the Euler equations in Fourier space.
For the mock catalogs we generated we used the Planck 2018 \cite{Planck2018} cosmological parameters. 

Using nbodykit averaging functions, we obtain the density and velocity fields on a discrete three-dimensional grid defined over a cube of edge length 300 Mpc, consisting of $11^3$ cubical cells of equal edge length.
The mock catalogs obtained in this way have statistical properties in good agreement with the results of more computationally expensive N-body simulations \cite{Hand:2017pqn}.
Examples of the simulated density and velocity fields are given in figs.(\ref{df}-\ref{vf}).

\begin{figure*}
    \includegraphics[scale=0.42,angle=90]{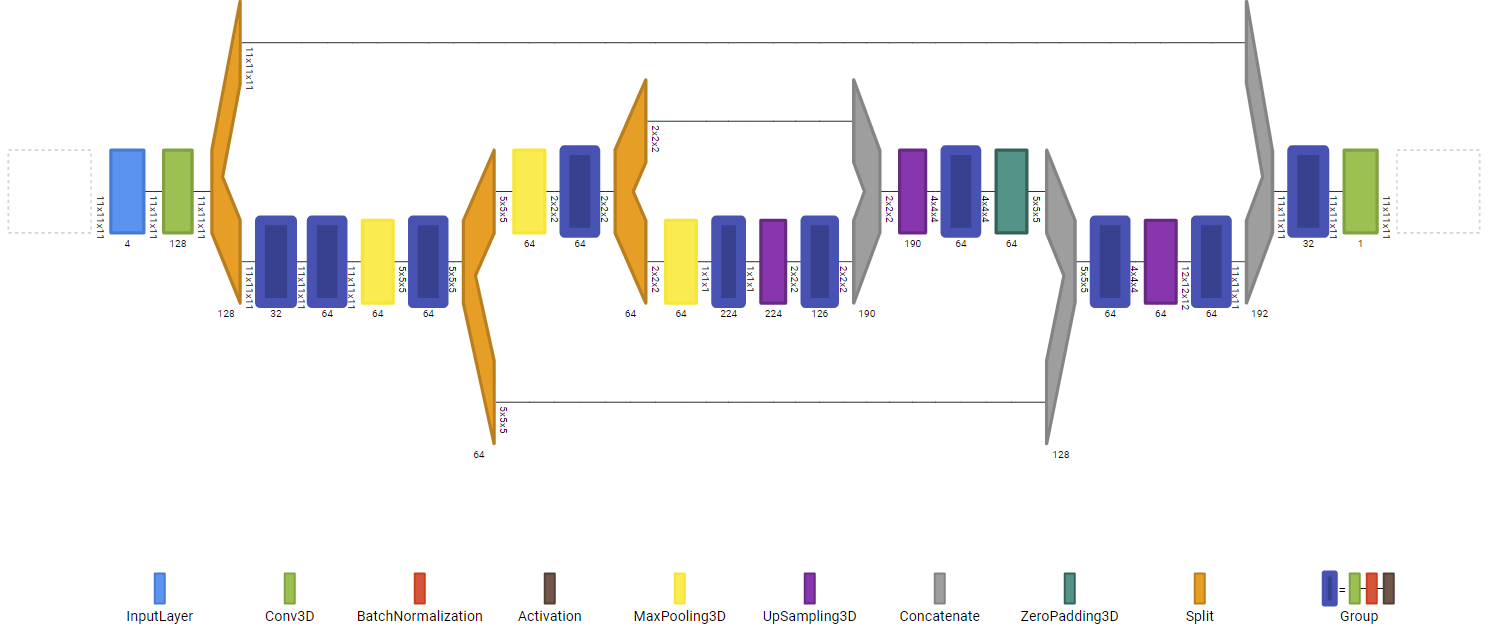}
    \caption{Diagram of the neural network used to solve the inversion problem, based on a modified version of U-Net. The colors represent different types of layers and the grey lines between the split and concatenate layers represent the residual connections. The number under the layers correspond to the number of channels of the layer.}
\end{figure*}


\section{Effects of cosmic structure on the luminosity Distance}
The luminosity distance is affected by inhomogeneities, and the effects can be found in  different relativist perturbative calculations  \cite{Misao,Bonvin_2006}.
There are different physical effects contributing to the difference with respect to the a background Friedman-Robertson-Walker (FRW) space, but at low redshift the most important ones are the Doppler effect due to the source and observer peculiar velocity, and the integrated lensing effect, while the effects associated to the time evolution of the gravitational potential along the line of sight can be neglected.

In the Newton gauge the dominant effect can be written as formula \cite{Bonvin_2006}
\be
D_L(z_S)=(1+z_S)(\chi_0-\chi_S)(1-k_v-k_\delta) \,, \label{DL}
\ee
where $z$ is the red-shift, $\chi$ is the comoving distance, and $k_v$ and $k_\delta$ correspond to the effects  \cite{Bolejko:2012uj} of the peculiar velocity and the density contrast

\be
k_v=\frac{1+z_S}{(\chi_O-\chi_S)H_s}\pmb{v_O}\cdot \pmb{n}+\left[1-\frac{1+z_S}{(\chi_O-\chi_S)H_s}\right]\pmb{v_S}\cdot \pmb{n},
\ee

\be
k_\delta=\frac{3}{2}H_0^2 \Omega_m \int_{\chi_S}^{\chi_O} \frac{\chi(\chi_S-\chi)[1+z(\chi)]}{(\eta_O-\eta_S)}\delta \, d\eta\,. \label{kd}
\ee
where $\chi$ is the comoving coordinate, and $\delta$ is the density contrast.

In the above equations subscripts ${}_s$  and  ${}_o$ denote quantities evaluated at the source and observer, $\pmb{n}$ is the unit vector between the source and the observer,  and $H_S=H(z_S)$ is the Hubble parameter at the source, obtained from the Friedmann's equation
\be
H(z)=H_0\sqrt{\Omega_\Lambda+\Omega_m (1+z)^3}\,.
\ee
The red-shift $z(\chi)$ associated to a given comoving distance is obtained by inverting numerically the relationship 
\begin{equation}
    \chi(z)=\int_0^z \frac{dz^\prime}{H(z^\prime)} \,.
\end{equation}{}
Using eq.(\ref{DL}) we obtain the luminosity distance  for each cell of the grid. For each simulation we generate a density profile grid $\delta_G$ with $L^3$ elements, a velocity grid $v_G$ with $L^3$  elements, corresponding to the line of sight projection of the source velocity $\pmb{v_S}\cdot \pmb{n}$, and a luminosity distance grid $D_G$ with $L^3$ elements.

The solution of the inversion problem can be summarized in this way:
\begin{itemize}
\item generate random grids for the radial velocity and density fields $\{\delta_G,v_G\}$
\item compute the  luminosity distance grid $D_G = f(\delta_G,v_G)$ for each $\{\delta_G,v_G\}$ using eq.(\ref{DL})
\item train a CNN to invert the above relationship i.e. to obtain $\{\delta_G,v_G\} = f^{-1}(D_G)$
\end{itemize}
where we are denoting symbolically with $f$ the solution of the direct problem, and with $f^{-1}$ the solution of the IP.
The solution of the IP is in general not unique, since different inhomogeneities could produce the same effects on $D_L$, especially due to the integrated effect in eq.(\ref{kd}). We train the CNN in order to minimize the error of the IP, within the limits of the above mentioned intrinsic degeneracy of the IP solution.

\section{Neural network architecture}
We solve the IP using convolutional neural networks, specifically a modified version of the U-Net architecture \cite{ronneberger2015unet}. We train separately two CNN, one for the density field $\delta_G$ and another for the velocity field $v_G$. The architecture is symmetrical and has two main parts: the left part is the encoder, which consists of 3D convolutional layers, while the right side is the decoder.
The latter uses nearest neighbor interpolation to upsample the data which is used to reconstruct the density and velocity fields from their effects on the luminosity distance. 

The simulated input data used to train the CNN is a four-dimensional array of  size $11*11*11*4$. This is because for each of the $11^3$ cells of the grid there is the corresponding value $D_G$, and the three spatial coordinates of the cell. This is necessary in order to provide information about the spatial location of the cells, which is important to determine the effects on the luminosity distance. The output of the CNNs  are   three-dimensional arrays of size $11*11*11$, which corresponds to the density and radial component of the source velocity in each of the grid cells.

\section{Loss function}
We train two different networks, one for $\delta_G$ and the other for $v_G$. In both cases we minimize the loss function given by the  Mean Absolute Error defined as
\begin{equation}
    MAE(y,\hat{y})=\frac{1}{m}\sum_{i=1}^m \frac{1}{L^3} \sum_{c=1}^{L^3}|y_{ic}-\hat{y}_{ic}|,
\end{equation}
where $m$ is the total number of simulations, $y_{ic}$ and $\hat{y}_{ic}$ are the inputted and predicted data respectively, for each cell $c$ in each simulation $i$. We apply early stopping to select the network parameters that best fit the test set in order to avoid overfitting and have good generalization results. 

Since both $\delta_G$ and $v_G$ can often have very small values, the Mean Average Percentage Error (MAPE) is not a good measure of the goodness of fit because it can often diverge, due to the presence of a small quantity in the denominator. For this reason we have chosen the MAE as loss function and metric to measure the inversion results accuracy.

\section{Results of the inversion of the density field}
\begin{figure}
    \includegraphics[scale=0.38]{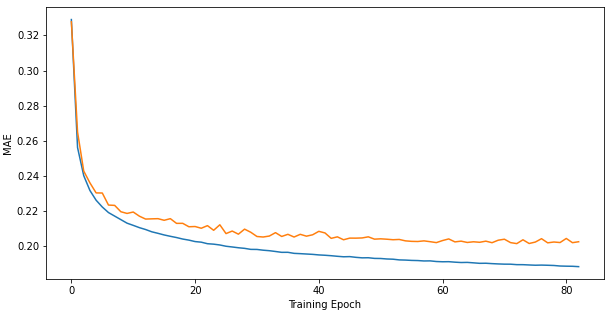}
    \caption{Training curves for the density contrast $\delta$. The orange and blue curves correspond respectively to the test and training datasets.}
    \label{mae_delta}
\end{figure}

\begin{figure}
    \includegraphics[scale=0.38]{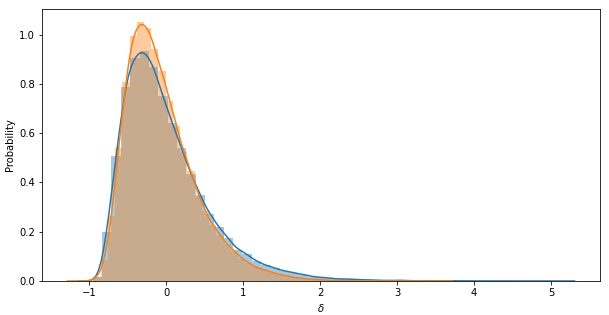}
    \caption{Probability distribution of the  $\delta$ field of the test set (blue), and corresponding  reconstructed $\delta$ (orange) obtained applying the neural network to the luminosity distance.}
    \label{dist_delta}
\end{figure}

\begin{figure}
    \includegraphics[scale=0.2]{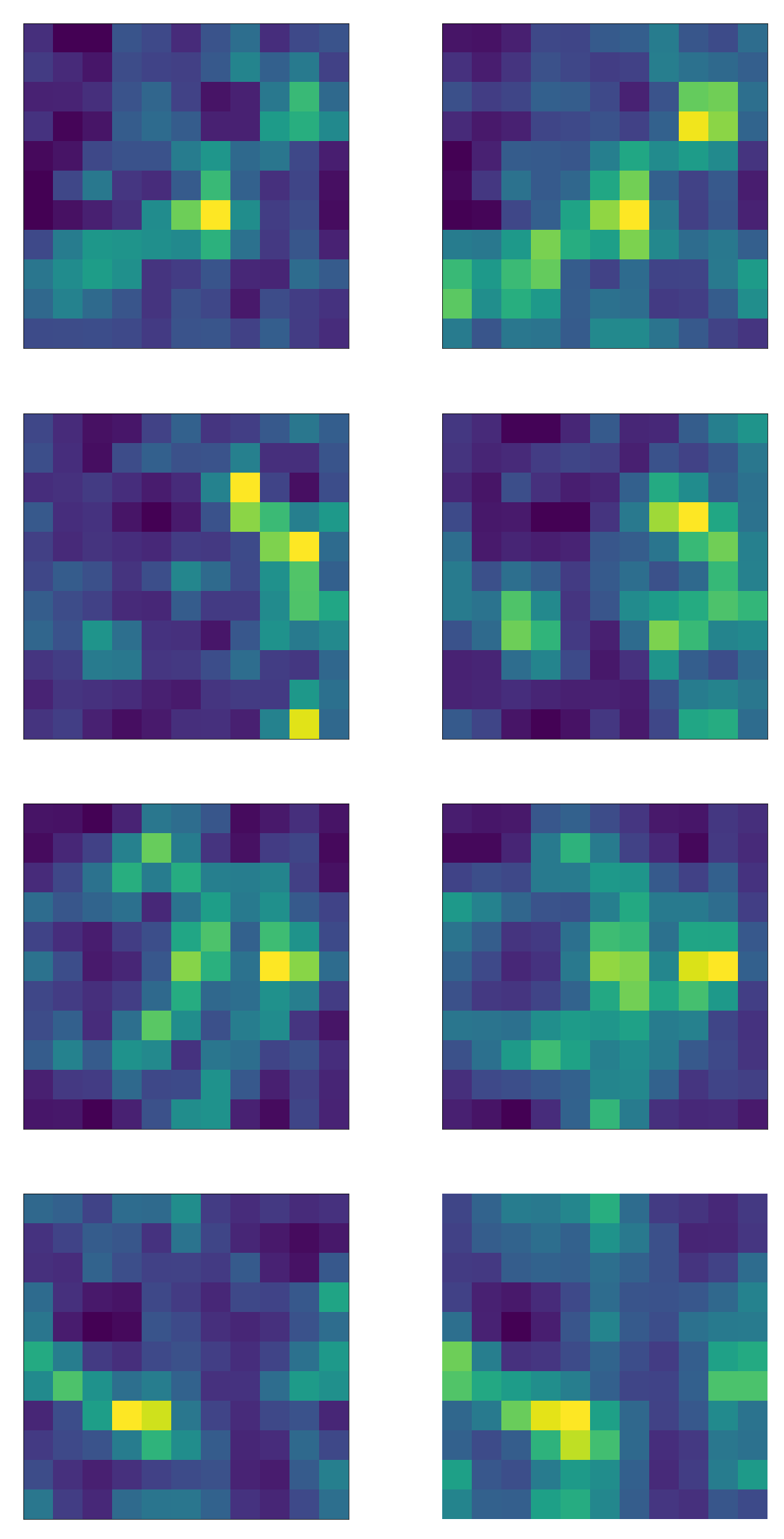}
    \caption{Different cross sections of the three-dimensional density contrast field. The left column is for the test set, and the right column is the corresponding $\delta$ reconstructed from the luminosity distance.}
    \label{cut_delta}
\end{figure}

The learning curves for $\delta$ are shown in fig. (\ref{mae_delta}). 
The probability distribution of the  $\delta$ field of the test set, and the corresponding  reconstructed $\delta$ obtained applying the neural network to the luminosity distance are shown in fig.(\ref{dist_delta}). As it can be seen  the reconstructed $\delta$ follows approximately the same distribution of the test set, showing that the neural network is able to recover the statistical properties of the test data set.
Different cross sections of the test set $\delta$ and their corresponding reconstructions are shown in fig. (\ref{cut_delta}), showing a qualitatively good agreement.
The MAE of the reconstructed $\delta$ is $0.20$, considerably smaller than the standard deviation $\sigma_{\delta}=0.55$ of the test set, which can be used as a benchmark to asses the accuracy of the inversion.


\section{Results of the inversion of the peculiar velocity field}
\begin{figure}
    \includegraphics[scale=0.38]{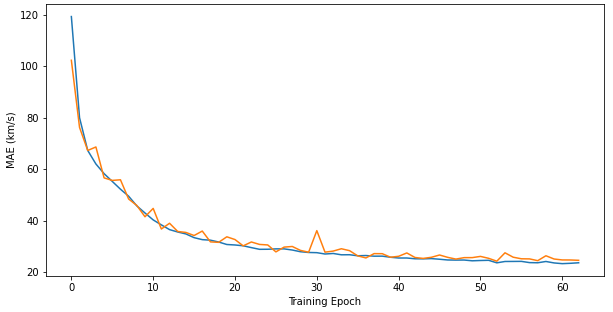}
    \caption{Training curve for the velocity. The training and test data sets correspond to the blue and orange lines respectively.}
    \label{mae_vsrc}
\end{figure}

\begin{figure}
    \includegraphics[scale=0.37]{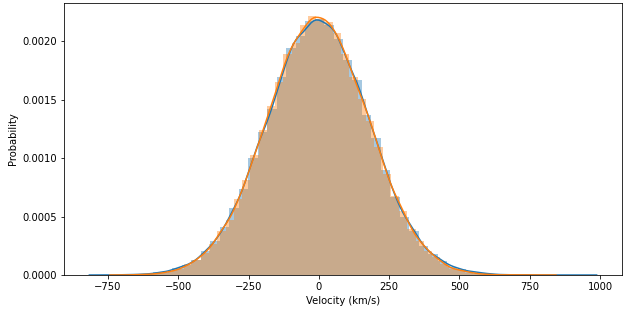}
    \caption{Probability distribution of the velocity of the test set(blue) and of the reconstructed velocity (orange).}
    \label{dist_vsrc}
\end{figure}

\begin{figure}
    \includegraphics[scale=0.2]{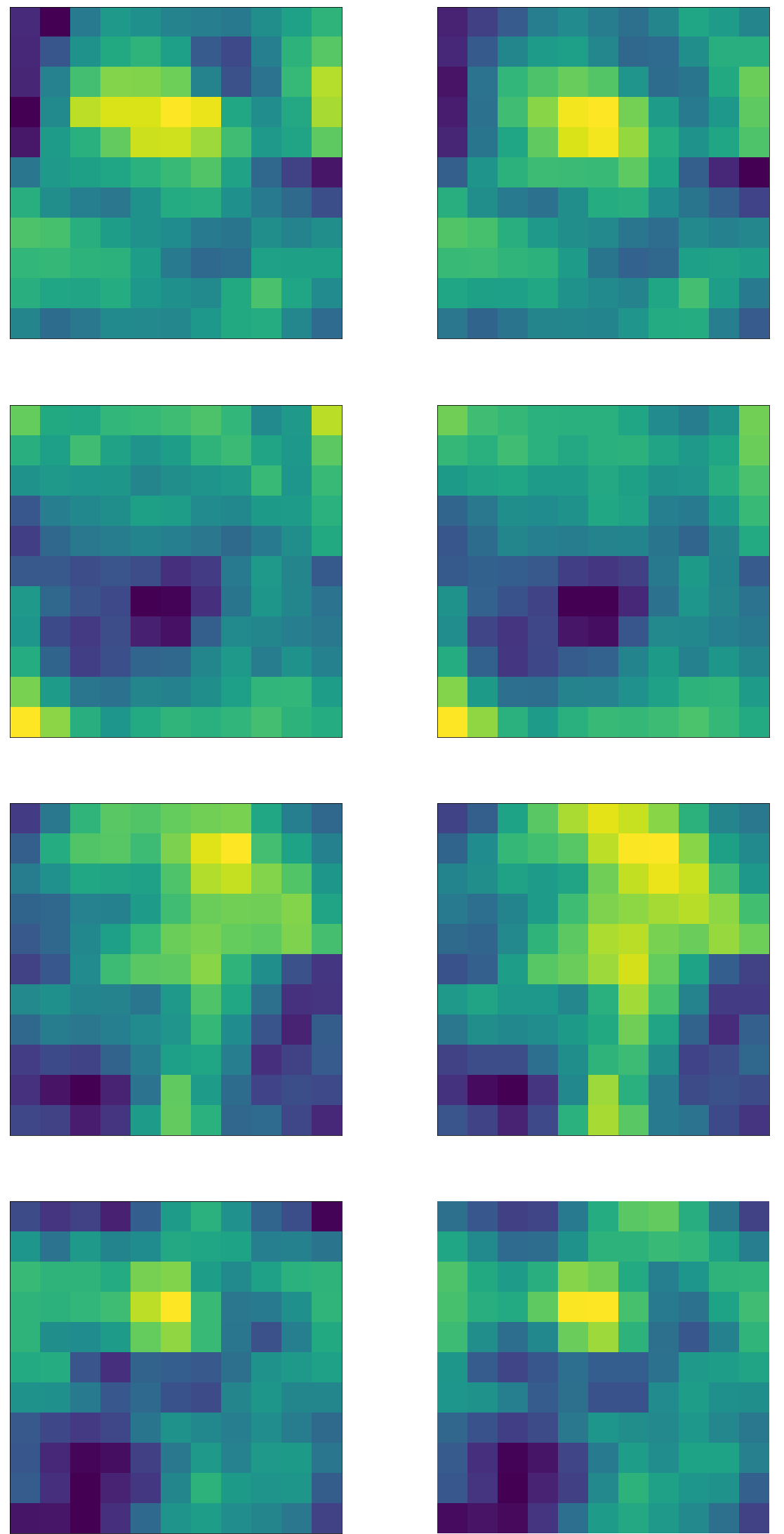}
    \caption{Cross section of the velocity field for different simulations in the test set. The left column corresponds to the velocity field of the test set and the images on the right column are their respective reconstructions.}
    \label{cut_vsrc}
\end{figure}

The mean absolute error of the reconstructed source velocity  is shown in fig.(\ref{mae_vsrc}). There is no apparent overfit between the training and test curves.
The probability distributions of the velocity of the test set and that of the reconstructed velocity plotted in fig. (\ref{dist_vsrc}) are in good agreement, showing that the neural network is able to recover the statistical properties of the test data set. 
An example of the cross sections the reconstructed tree-dimensional velocity field is shown in fig. (\ref{cut_vsrc}).
The MAE of the reconstruction is $24.33\, km/s$ which is quite smaller than the standard deviation of the velocity of the test set $\sigma_v=174.82$ km/s, showing a good performance of the neural network.
\section{Why the velocity field is reconstructed better than the density?}

As shown in the previous sections the reconstruction of the velocity field is more precise than that of the density field. We trained the two networks separately, using different training sets, one for the velocity and the other for the density, with the goal to determine which field can be reconstructed more accurately.

The difference in the reconstruction accuracy is due to the fact that density field has an integrated effect on the luminosity distance as shown in eq.(\ref{kd}), implying that different line of sight density field configuration can produce the same effect. This degeneracy in the solution of the inversion problem, causing  the lower accuracy of the reconstructed density profiles, is not specific to this case, but is the manifestation of the mathematical properties of this kind of problems, leading to the non-uniqueness of the solution of the inversion problem, as well known in other contexts such as for example in seismic inversion \cite{1986InvPr...2L..23R,10.1093/gji/ggab134}.
While the non-uniqueness of the solution of the inversion problem implies an  intrinsic uncertainty on the accuracy of the results, the reconstructed fields still provide valuable information which at high redshift cannot be obtained in any other way.

In the case of the radial source velocity instead, this is a single scalar quantity for each cell of the grid, and its effect on the luminosity distance only depends on it, not on any other quantity along the line of sight.
This implies that there is a unique physical configuration which can produce a given effect on the luminosity distance, and consequently no degeneracy in the solution of the inversion problem, and a higher accuracy of the reconstructed velocity field.

\section{Improving accuracy with density and velocity fields combined inversion}
In the future it will be interesting to design a new network architecture to be trained on the combined dataset of velocity and density fields, to see if this can improve the accuracy obtained when training separate networks on separate datasets.

The combined data set is expected to provide extra information to be learned by the network, since velocity and density fields are related to each other by the Euler's equations, and this should improve the accuracy of the reconstructed results. In fact, while the density field is more difficult to reconstruct due to its integrated effect on the luminosity distance, it is not independent from the velocity field, and the knowledge of latter should improve also the reconstruction of the density, partially reducing the degeneracy of the solution.

\section{Application to real data}
The training datasets used in this paper assumed a homogeneous distribution of the sources, i.e. one source was located in the center of each cell, since the main goal was to introduce this new approach to the inversion problem and show how it performs  on simulated data.
Standard candles are not uniformly distributed, and for this reason in order to train a network to apply it to observational data for inference, the distribution of sources in the training set should mimic, at least statistically, the observed distribution of SNe.

A solution would be training the neural network to handle the sources inhomogeneous distribution. This could be achieved by adding a dropout layer which masks random cells in the input data, so that the neural network would learn how reconstruct the density and speed fields from a masked luminosity distance grid, with some cells with no source. This dropout layer can also reduce the overfit, as the network would be able to generalize better when applied to the validation set or masked data. 

During training the amount of masked data should be increased gradually, from a totally unmasked luminosity distance grid during the first learning epochs, to a partially masked input grid, where the proportion of unmasked cells corresponds to the proportion of cells hosting observed SNe. This kind of modified neural network could then be used to perform more effectively inference of the density and velocity fields from SNe catalogues \cite{Pan-STARRS1:2017jku}, and we leave this interesting task to a future work.


\section{Conclusion}
We have obtained for the first time solutions of the IP problem for arbitrary space-time geometries, based on the use of convolutional neural networks to reconstruct the velocity and density fields from the luminosity distance of SNe.
The data to train the CNN is obtained by computing the effects on the luminosity distance of the density and velocity fields from numerical simulations of cosmic structure.

We trained a modified U-Net architecture which uses residual connections at three different scales. This fully convolutional network has roughly half a million parameters and is made up of 11 convolutional layers, and nearest neighbour interpolation is used to upsample the data in the decoder. The inputs of the encoder are the values of the luminosity distances for each cell in the grid and the coordinates of the cell. We used heldout validation and the network was trained for 60 and 80 epochs for the velocity and density fields respectively.

The CNN is able to reconstruct the test set density and velocity field with a mean average error of respectively $0.20$ and $24.33$ km/s, compared to a standard deviation of the test set of $\sigma_{\delta}=0.55$ and $\sigma_v=174.82$ km/s. The density field is more difficult to reconstruct due to its integrated effect on  the luminosity distance, implying that multiple density configurations along the line of sight can produce the same integrated effect, making the solution of the IP not unique.
The probability distributions of the reconstructed density and velocity fields match well the test set distribution. 

In the future it would be important to increase the resolution of the numerical simulations used to train the network, to increase the accuracy of the results of the inversion. It will also be interesting to apply this CNN to observational data. This task could be achieved  by adding a dropout layer to the network, so that the neural network can learn to reconstruct the density and velocity fields from non uniformly distributed SNe catalogues.
\section*{Data Availability}

The data underlying this article will be shared on reasonable request to the corresponding author.

\section*{Acknowledgements}
We thank the authors of nbodykit \cite{Hand:2017pqn} for interesting discussions about the use of the package.
A.E.R. is supported by the UDEA projects 2021-44670-UFS, 2019-28270-MGT, ES84190101, and by NAWA.
We also thank the anonymous Referee for the useful comments and suggestions.



\bibliographystyle{mnras}
\bibliography{example} 








\bsp	
\label{lastpage}
\end{document}